\documentclass{elsart}
 \usepackage{graphicx}

\usepackage{amssymb}

\newcommand{\eg}{{\it e.g.\ }}
\newcommand{\ie}{{\it i.e.\ }}
\renewcommand{\etal}{{\it et al.\ }}

\begin{document}

\begin{frontmatter}


\title{When are networks truly modular?}
\author{J\"org Reichardt and Stefan Bornholdt}
\ead{\{reichardt,bornholdt\}@itp.uni-bremen.de}

\address{Institute for Theoretical Physics, University of Bremen, Otto-Hahn-Allee, 28359 Bremen, Germany}


\begin{abstract}
Study of the cluster- or community structure of complex networks makes an important contribution to the understanding of networks at a functional level. Despite the many efforts, no definition of community has been agreed on and important aspects such as the statistical significance and theoretical limits of community detection are not well understood. We show how the problem of community detection can be mapped onto finding the ground state of an infinite range spin glass. The ground state energy then corresponds directly to the quality of the partition. The network modularity $Q$ previously defined by Girvan and Newman \cite{Girvan03} turns out to be a special case of this spin glass energy. Through this spin glass analogy, we are able to give expectation values for the modularity of random graphs that can be used in the assessment of the statistical significance of real network clusterings. Further, it allows for assessing the theoretical limits of community detection. 
\end{abstract}

\end{frontmatter}

\section{Introduction}
With the increasing availability and steadily increasing size of relational data-sets or networks the need for appropriate methods for exploratory data analysis arises. For general statistical properties such as the degree distribution, degree correlations, clustering etc. a number of well established methods and models to explain their origin exist \cite{BAReview,NewmanReview}. However, a standard analysis for the higher order structure in graphs has not been established so far. Currently, the problem of the cluster or community structure is subject of intense study \cite{Newman,GuileraReview}. Cluster analysis is an important technique that allows for data abstraction, dimensionality reduction or aids in data visualization.  It is used in life sciences \cite{GuimeraMeta}, over bibliometrics \cite{Palla05}, to market research \cite{ReichardtEbay}, and has implications for experiment planning, funding policies or marketing. 

However, some important aspects of the problem are still not clearly understood. First, there is little agreement with regard to the definition of a community or cluster in a network and second, and more importantly, no adequate measure of the statistical significance of network clustering exists. This article aims at contributing to both of these questions.     

\section{What is a community?}
Despite the many applications of community detection across the sciences, it remains remarkably unclear what a community actually is. Additionally to the many definitions that are given in sociology \cite{WassermanFaust}, the physics community has contributed a fair number as well \cite{Newman,GuileraReview}. All authors agree that a community should be a group of nodes that is more densely connected among each other than with the rest of network, but differ largely in the details. Below, we give a short overview of the different aspects that have been emphasized by different authors. 

The initial work on communities by Girvan and Newman \cite{Girvan} gives an algorithmic definition. They design a community detection algorithm which recursively partitions the graph to produce a hierarchy of communities from the entire network down to single nodes. At each point, the nodes belonging to distinct sub-trees in the resulting dendogram are considered as communities.   

Radicchi \etal \cite{Radicchi} tried to improve this heuristic definition by coining the term of ``community in a strong sense'' such that
\begin{equation}
k_i^{in}>k_i^{out}, \forall i\in C.
\end{equation}
This means for all nodes $i$ in the community $C$, the number of connections node $i$ has to members of its own community $k_i^{in}$ is larger than $k_i^{out}$, the number of connections is has to the rest of the network. Further, they define a ``community in a weak sense'', such that the sum of internal connections is larger than the sum of external links $\sum_{i\in C}k_i^{in}>\sum_{i\in C}k_i^{out}$. Radicchi \etal then suggest to stop any recursive partitioning algorithm when an additional partition would not result in a community in the strong (or weak) sense.  

Palla \etal \cite{Palla05,PallaPRL} have given a definition based on reachability. They define a sub-graph percolation process based on k-cliques (fully connected subgraphs with $k$ nodes). Two k-cliques are connected, if they share a (k-1)-clique, \eg two triangles (which are 3-cliques) are connected if the share an edge (a 2-clique). A community, or k-clique percolation cluster, is then defined as the group of nodes that can be reached via adjacent k-cliques. Communities may overlap, \ie nodes may belong to more than one percolation cluster, but communities corresponding to (k+1)-clique percolation clusters always lie completely within k-clique clusters.   

Girvan and Newman have further defined a quantitative measure of the quality of an assignment of nodes into communities.  This so-called ``modularity'' \cite{Girvan03} can be used to compare different assignments of nodes into communities quantitatively. The modularity is defined as:
\begin{equation}
Q=\sum_se_{ss}-a_s^2.
\end{equation} 
The sum runs over all communities $s$. The fraction of all links connecting nodes in group $s$ and $r$ is denoted by $e_{sr}$. Hence, $e_{ss}$ is the fraction of all links lying within group $s$. The fraction of all links connecting to nodes in group $s$ is denoted by $a_s=\sum_re_{rs}$. One can interpret $a_s^2$ as the expected fraction of internal links in group $s$, if the network was random and the nodes were distributed randomly into the different groups. Such a measure can be used to stop recursive partitioning or agglomerative approaches when they do not lead to an improvement of $Q$ anymore \cite{FastGN}.  

We see the diversity of definitions and approaches of which we have described only a few. References \cite{Newman,GuileraReview} give a more comprehensive overview. Because of this controversy of opinions, we will set out from a first principles approach in the next section that will shed some light on the general properties of the problem.

\section{A first principles approach to community detection}
Instead of defining what a community is and then trying to devise an algorithm in order to detect it, we use a different approach. We start from a simple principle: to group nodes that are not linked in different communities and to put nodes which are linked in the same community. With this principle, we write the following Hamiltonian:
\begin{equation}
\mathcal{H}(\sigma)=-\sum_{i<j}a_{ij}A_{ij}\delta(\sigma_i,\sigma_j)+\sum_{i<j}b_{ij}(1-A_{ij})\delta(\sigma_i,\sigma_j).
\label{Ham1}
\end{equation}
Here, $\sigma_i$ denotes the group index of node $i$, $\delta(\sigma_i,\sigma_j)$ is the Kronecker delta, $A_{ij}$ is the adjacency matrix of the network with $A_{ij}=1$ if nodes $i$ and $j$ are connected and zero otherwise. Hence, the first sum runs over all pairs of connected nodes, while the second sum runs over all pairs of unconnected nodes. Our Hamiltonian rewards every pair of connected nodes $(i,j)$ in the same group with $a_{ij}$ and penalizes every pair of unconnected nodes $(i,j)$ in the same community with $b_{ij}$. It implements just the principle we started out from. Any spin configuration that will minimize (\ref{Ham1}) is hence optimal in the sense of this first principle. It is now important to define the weights $a_{ij}$ and $b_{ij}$ in a sensible way. A particular good choice is to balance them, such that all existing connections in the network are equally important to our optimality criterion as all missing connections, of which there are generally many more: 
\begin{equation}
\sum_{i<j}a_{ij}A_{ij}=\sum_{i<j}b_{ij}(1-A_{ij}).
\label{WeightRelation}
\end{equation} 
One way of satisfying this equation is to set $a_{ij}=1-\gamma p_{ij}$ and $b_{ij}=\gamma p_{ij}$ which also reduces the need for two different weights to only one. We have introduced an additional parameter $\gamma$ that will allow us to adjust the balance of missing and existing links. The only constraint we have to impose on $p_{ij}$ in order to fulfill (\ref{WeightRelation}) is that $\sum_{i<j}p_{ij}=M$ with $M$ being the total number of links in the network. With this choice, we can now rewrite (\ref{Ham1}) in a much simpler form:
\begin{equation}
\mathcal{H}(\sigma)=-\sum_{i<j}\left(A_{ij}-\gamma p_{ij}\right)\delta(\sigma_i,\sigma_j).
\label{Ham2}
\end{equation}
Equation (\ref{Ham2}) is formally identical to the Hamiltonian for a $q$-state Potts spin glass, with $q$ being the number of possible group indices. The coupling matrix is then defined as $J_{ij}=A_{ij}-\gamma p_{ij}$. 
Though $p_{ij}$ can take any form, it is sensible to identify it with the connection probability between nodes $i$ and $j$ in the network. Depending on the network under study, this can be 
\begin{equation}
p_{ij}=p, 
\end{equation}
if the links are assumed to connect nodes with constant probability $p=2M/N(N-1)$. Another possible choice is 
\begin{equation}
p_{ij}=\frac{k_ik_j}{2M},
 \label{Uncorrelated}
\end{equation} 
if the degree distribution of the nodes is to be taken into account and there are no degree-degree correlations. Here $k_i$ denotes the degree of node $i$ and $M$ represents the number of links in the network as before.   
 
Both of these choices render $p_{ij}$ positive and smaller than one, hence we are dealing with a spin glass which has ferromagnetic couplings between connected node and anti-ferromagnetic couplings between unconnected nodes. The ground state of this spin glass defines the optimal assignment of nodes into communities. Note, that for $\gamma=1$ and $p_{ij}=k_ik_j/2M$, we recover the modularity  $Q$ defined by Girvan and Newman \cite{Girvan03} from (\ref{Ham2}) via $Q=-\frac{1}{M}\mathcal{H}$ \cite{NewmanLarge}.    

It is worth rewriting (\ref{Ham2}) as a sum over spin states $s$: 
\begin{equation}
\mathcal{H}=-\sum_s\underbrace{\left(m_{ss}-\gamma[m_{ss}]_{p_{ij}}\right)}_{c_{ss}}=\sum_{s<r}\underbrace{\left(m_{rs}-\gamma[m_{ss}]_{p_{ij}}\right)}_{a_{rs}}.
\label{Modularity2}
\end{equation}
We denote the number of links within group $s$ by $m_{ss}$ and between groups $r$ and $s$ by $m_{rs}$. Further, we denote the expectation values of these quantities under the model of connection probability $p_{ij}$ and assuming a random assignment of spins by $[\cdot]_{p_{ij}}$. In (\ref{Modularity2}) we have introduced two new terms $c_{ss}$ and $a_{rs}$ which measure within group ``cohesion'' and between group ``adhesion'', respectively. Note that maximizing cohesion and minimizing adhesion are in fact equivalent and will hence always be extremal at the same time, \ie any configuration of spins that minimizes $\mathcal{H}$ will automatically maximize cohesion and minimize adhesion.     

In particular, we have for the two connection models introduced above
\begin{equation}
[m_{ss}]_{p_{ij}}=p\frac{n_s(n_s-1)}{2},\hspace{1cm}\mbox{and}\hspace{1cm}[m_{rs}]_{p_{ij}}=pn_sn_r
\end{equation}
for $p_{ij}=p$. The number of nodes in group $s$ is denoted by the occupation number $n_s$ \cite{ReichardtPRL}. For $p_{ij}=k_ik_j/2M$ we find
\begin{equation}
[m_{ss}]_{p_{ij}}=\frac{1}{2M}\frac{K_s^2}{2},\hspace{1cm}\mbox{and}\hspace{1cm}[m_{rs}]_{p_{ij}}=\frac{1}{2M}K_sK_r
\end{equation}
where $K_s$ denotes the sum of degrees of nodes in group $s$ in a similar way as the occupation number $n_s$. 

We have thus shown, that finding the community structure of a network is equivalent to finding the ground state of an infinite range spin glass. Note that non-zero couplings exist between all pairs of nodes. Fortunately, the particular choice of $p_{ij}$ allows us to implement efficient optimization routines \cite{ReichardtPRE} that only need to consider interactions along the links and treat the anti-ferromagnetic interactions along the non existing links in a mean field manner, which is, however, not an approximation but accounts exactly for the repulsive interactions. One only needs to keep track of the occupation numbers of the spin states or the total sum of degrees in each group.  

We are now able to give a definition of community which follows directly from the properties of the ground state as a global minimum of (\ref{Ham2}). We define as \textit{the} community structure of a graph an assignment of nodes into groups (spin states) that makes (\ref{Ham2}) minimal. Such an assignment then possesses the following properties:
\begin{enumerate}
 \item Every proper subset $n_1$ of a community $n_s$ has a maximum coefficient of adhesion with its complement in the community compared to the coefficient of adhesion with any other community ($a_{1,s\backslash 1}=\mbox{max}$).  
\item The coefficient of cohesion is non-negative for all communities ($c_{ss}\geq 0$).
\item The coefficient of adhesion between any two communities is non-positive ($a_{rs}\leq 0$).  
\end{enumerate}
This also defines the term ``community''. A community is a group of nodes that has the above three properties. 

If the ground state is degenerate, \ie different assignments of nodes into communities lead to the same ground state energy, this allows us to define overlapping community structure in a natural way. Degeneracy may occur in two different forms. On one hand it may be possible to move part of a community $a$ to another community $b$ without increasing the energy. We say the two communities $a$ and $b$ overlap, since the total number of communities stays constant. On the other hand, it may be that one may split a community $a$ into two or more communities or join it with another community $b$ without increasing the energy. Since the number of communities changes, we speak of overlapping community structures. Naturally, all groups of nodes with equal spin value in any configuration that represents a local minimum of (\ref{Ham2}) will also qualify as communities. They can be regarded as sub-optimal assignments and the study of their overlap among each other and with the ground state yields valuable information about how many alternative, but sensible groupings exist for a particular network \cite{ReichardtPRL,ReichardtPRE}  

The ground state depends on the value $\gamma$ chosen. The value of $\gamma$ at which the community structure was obtained should always be quoted. Changing the value of $\gamma$ allows to detect hierarchies in the assignment of nodes into communities \cite{ReichardtPRL,ReichardtPRE}. 

We benchmarked the performance of this approach to community detection on computer generated test networks \cite{ReichardtPRL} and compared the results to those obtained by Girvan and Newman's  betweenness algorithm \cite{Girvan}. The networks are Er\H{o}s-R\'{e}nyi (ER) graphs \cite{Erdos} with an average degree of $\langle k\rangle=16$ and $128$ nodes. The nodes were divided into 4 groups of 32 nodes each. Keeping the average degree fixed, the links per node were distributed into and average of $\langle k_{in}\rangle$ to members of the same group and an average of   $\langle k_{out}\rangle$ to members of the 3 remaining groups in the network such that $\langle k_{out}\rangle+\langle k_{in}\rangle=\langle k\rangle$. Obviously, increasing $\langle k_{out}\rangle$ on the expense of  $\langle k_{in}\rangle$ makes the recovery of the designed community structure more difficult. At $\langle k_{in}\rangle=4$ the network should be completely random and any trace of the built-in community structure is lost since at this point the probability to link to a member of a different node equals the probability to link to a member of the same group $p_{in}=p_{out}=p$.  

\begin{figure}
\includegraphics[width=7cm]{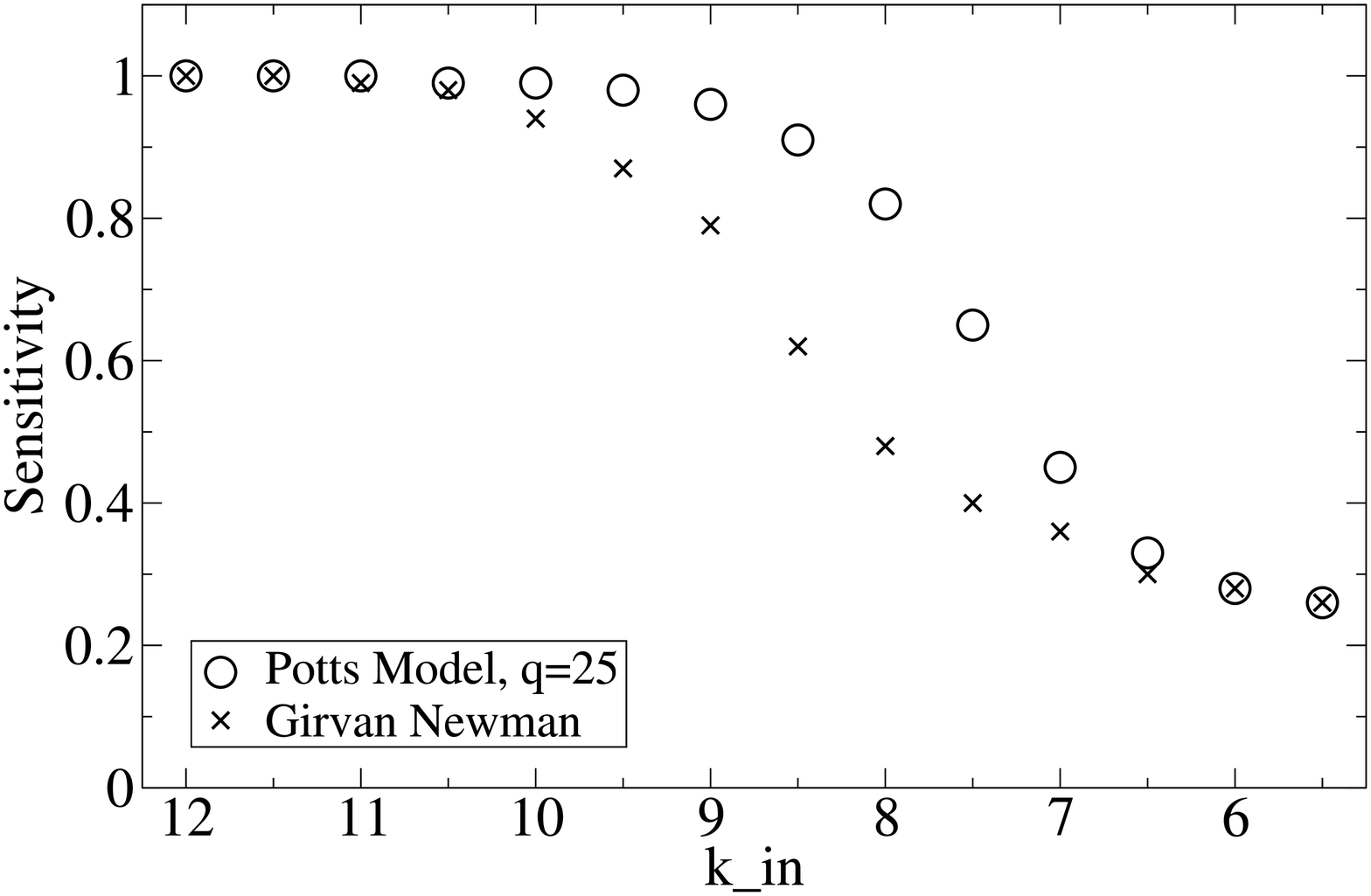}
\includegraphics[width=7cm]{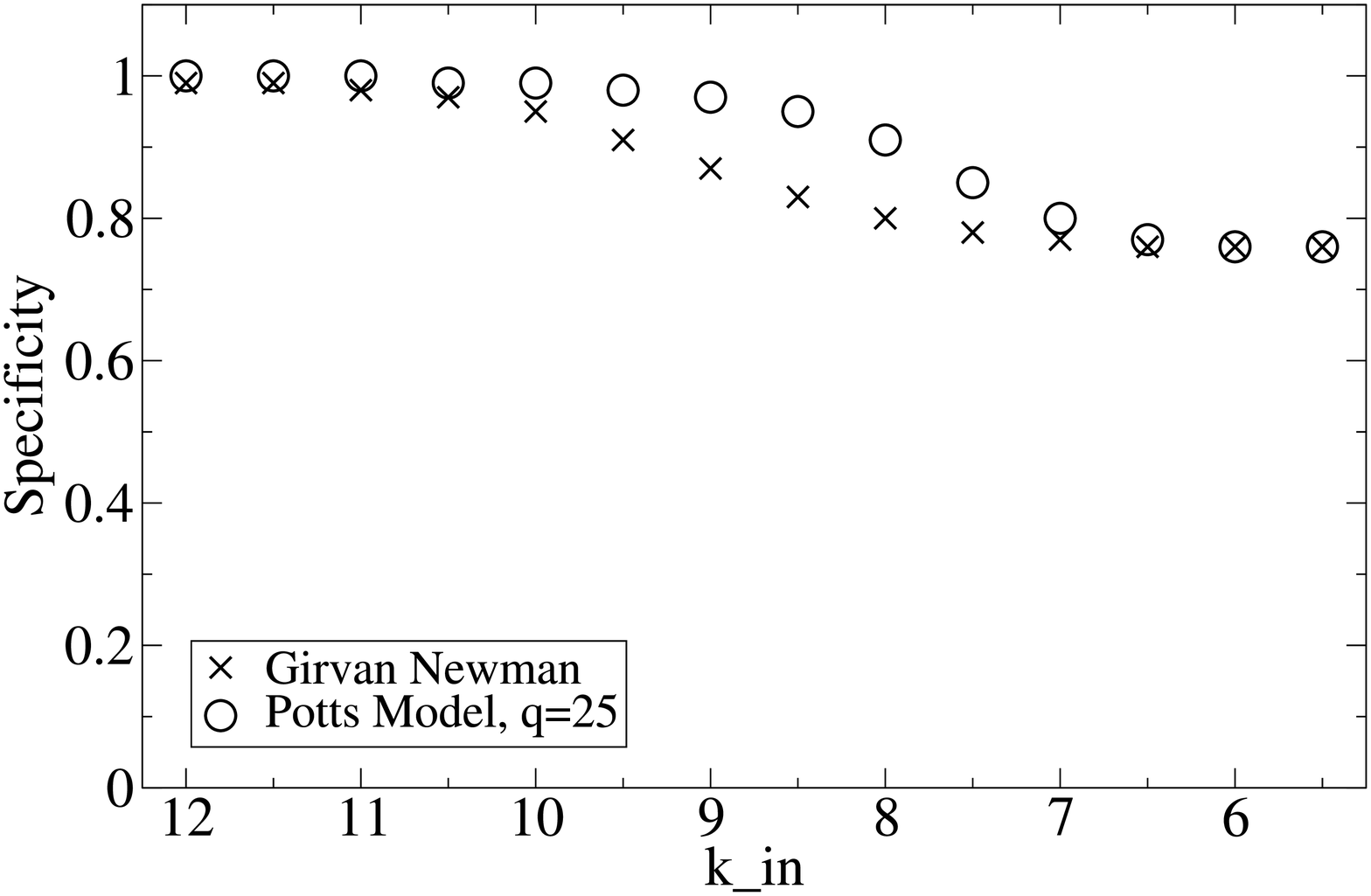}
\caption{Benchmarks of a community detection algorithms based on finding the ground state of a spin glass and comparison with Girvan-Newman's Algorithm \cite{Girvan}. Tests were run on computer generated test networks with known community structure. ``Sensitivity'' denotes the fraction of all pairs of nodes that are classified correctly in the same community, while ``specificity'' denotes the fraction of all pairs of nodes classified correctly in different communities.}
\label{Benchmark}
\end{figure}

Figure \ref{Benchmark} shows the results of the benchmarks. We measured the success of the two methods in terms of sensitivity and specificity. Sensitivity measures the fraction of pairs of nodes that are correctly classified as being in the same cluster, while specificity measures the fraction of nodes correctly classified as belonging to different clusters. In other words, the two measures indicate how good the algorithms are in grouping together what belongs together and in keeping apart what does not belong together. From Figure \ref{Benchmark} we see, that both algorithms are rather conservative in terms of grouping things together as indicated by the high levels of specificity. The change in sensitivity is much more drastic and we find that the Potts model approach outperforms the algorithm of Girvan and Newman \cite{Girvan}. The critical value of $\langle k_{in}\rangle_c$ at which the ability to recover the built in community structure seems to be $\langle k_{in}\rangle_c=8$.
\section{Communities and Modularities in Random Networks}
In our introductory paragraphs, we have already raised the question when one may call a network truly modular. Obviously, running a clustering algorithm over a set of randomly generated data points will always produce clusters which, however, have little meaning. Similarly, minimizing the modularity Hamiltonian on a random graph results in a community structure which has all the desired properties. This does, of course, not mean that the graph we studied was in fact modular. A differentiation between graphs which are truly modular and those which are not can hence only be made if we gain an understanding of the intrinsic modularity of random graphs. By comparing the modularity of random graphs with that of real world graphs, we can assess whether the graphs under study are truly modular. 

Such a comparison can of course always be made by randomizing the network under study keeping the degree distribution invariant. Such algorithms then remove all correlations and community structures possibly present in the data. Comparing the results of clustering the empirical data and a randomized version of it can always give a clue to what extent the data shows modularity above that of a random network with the same degree distribution. Nevertheless, such analysis is biased by the algorithm used to detect the community structure. Much more desirable would be a measure of modularity that can be used to compare with any algorithm. 

In mapping the problem of finding a community structure onto finding the ground state of an infinite range spin glass, we have defined a coupling matrix $J_{ij}$ with the following distribution of couplings:
\begin{equation}
p(J_{ij})=p_{ij}\delta(J_{ij}-(1-p_{ij}))+(1-p_{ij})\delta(J_{ij}+p_{ij}),
\end{equation} 
where we have set $\gamma=1$ and assumed we are dealing with a random network in which the links are distributed with the same $p_{ij}$ we use for defining the weights $a_{ij}$ and $b_{ij}$ of the contributions of existing and missing links in the clustering. It is easy to see that this distribution has zero mean. Since the mean of the distribution of couplings couples only to the magnetization, we find a zero magnetization in the ground state \cite{FuAnd}. This corresponds to an equi-partition of the network. The community structure of a random network consists of all equal sized communities. A symmetry argument can also be invoked to understand this. In an uncorrelated random graph, there is no reason for a particular size of communities and hence, they must be of equal size. If we conceive community detection as looking for the ``natural partition'' of a network, then the natural partition of a random graph is the equi-partition.

For the number of edges to cut when equi-partitioning a random graph, a number of results exist since the 1980's, beginning with the paper by Fu and Anderson \cite{FuAnd} about bi-partitioning a random graph. Kanter and Sompolinsky \cite{KanterSompQPart} have given an expression for the minimum total number of inter community edges $\mathcal{C}$, also called cut-size, when partitioning a random graph into $q$ equal sized parts: 
\begin{equation}
 	\min_{\{\sigma_i\}}\left[\mathcal{C}(\{N/q\})\right]=\frac{N^2p(q-1)}{2q}-N^{3/2}J\frac{U(q)}{q}.
 \end{equation} 
The minimum is taken over all possible spin configurations $\{\sigma\}$ with equal occupation numbers $n_s=N/q$. The first term in the expression is the expectation value of $\mathcal{C}$ for a random assignment of spin states and the second term is a correction due to optimization of the configuration which depends on the standard deviation of the coupling matrix $J$ and a constant depending on the number of parts $q$. In case of an ER-graph, the standard deviation of the coupling matrix is given by given by $J=\sqrt{p(1-p)}$ with $p$ denoting the average connection probability in the network given by $p=2M/N(N-1)$. From this, we can immediately write an expectation value for the modularity of random graphs:     
\begin{equation}
Q=-\frac{1}{M}\mathcal{H}_{GS}=\frac{N^{3/2}}{M}\sqrt{p(1-p)}\frac{U(q)}{q}.
\label{PottsModularity}
\end{equation}
For the $U(q)$, the ground state energy of a q-state Potts model with Gaussian couplings of zero mean and variance $J^2$, some values for small $q$ are given in Table \ref{PottsGS} obtained by using the exact formula for calculating $U(q)$ from \cite{KanterSompQPart}. For large $q$, we can approximate $U(q)=\sqrt{q\ln q}$ \cite{KanterSompQPart}.
 
\begin{table}
\begin{tabular}{|l||c|c|c|c|c|c|c|c|c|}
\hline
$q$ & $2$ & $3$  & $4$ & $5$ & $6$ & $7$ & $8$ & $9$ & $10$\\
$U(q)/q$ & $0.384$ & $0.464$ & $0.484$ & $0.485$ & $0.479$ & $0.471$ & $0.461$ & $0.452$ & $0.442$\\
\hline
\end{tabular}
\caption{Values of $U(q)/q$ for various values of $q$ obtained from  \cite{KanterSompQPart}, which can be used to approximate the expected modularity with equation (\ref{PottsModularity}).}
\label{PottsGS}
\end{table}

\begin{figure}
\includegraphics[width=7cm]{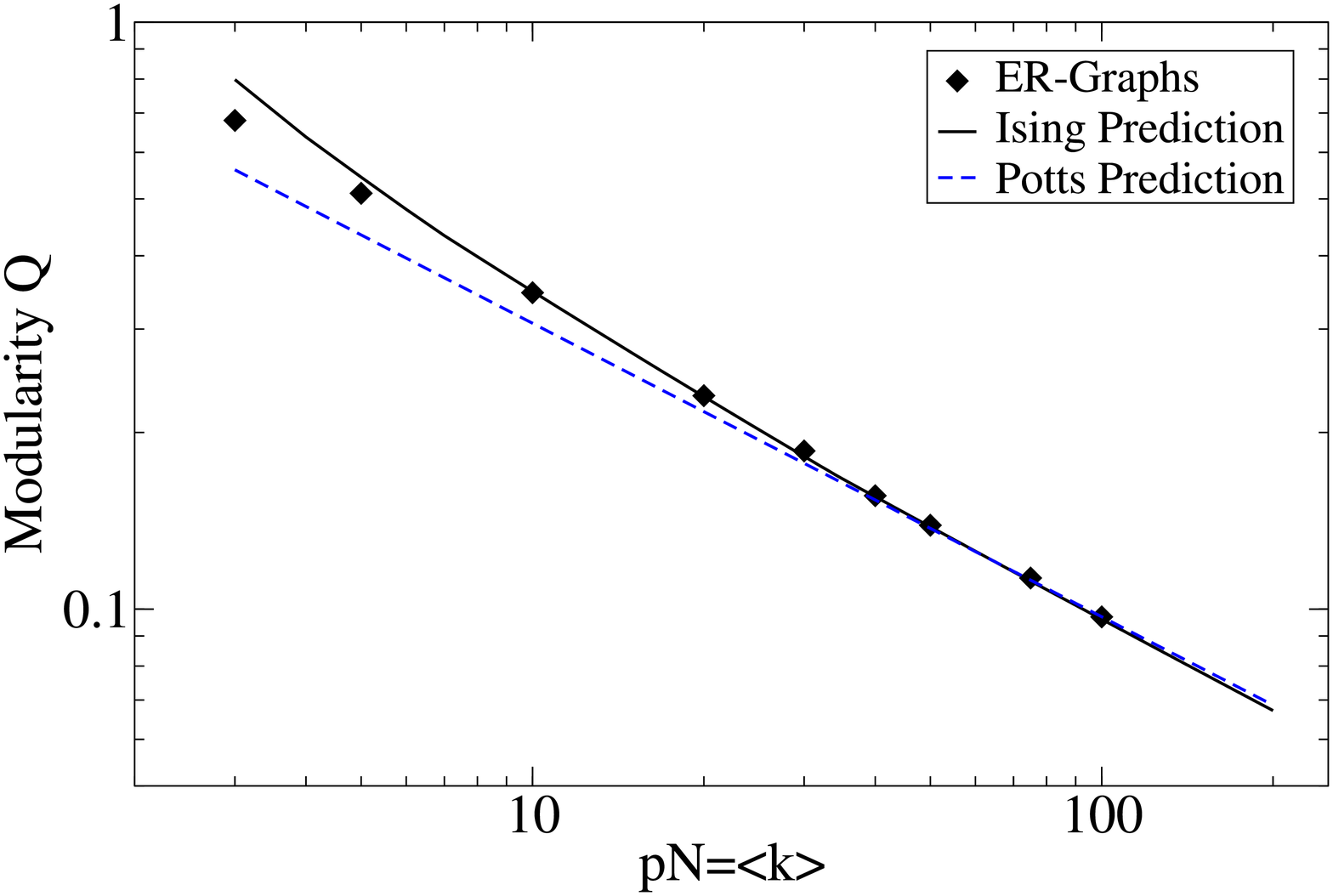}
\includegraphics[width=7cm]{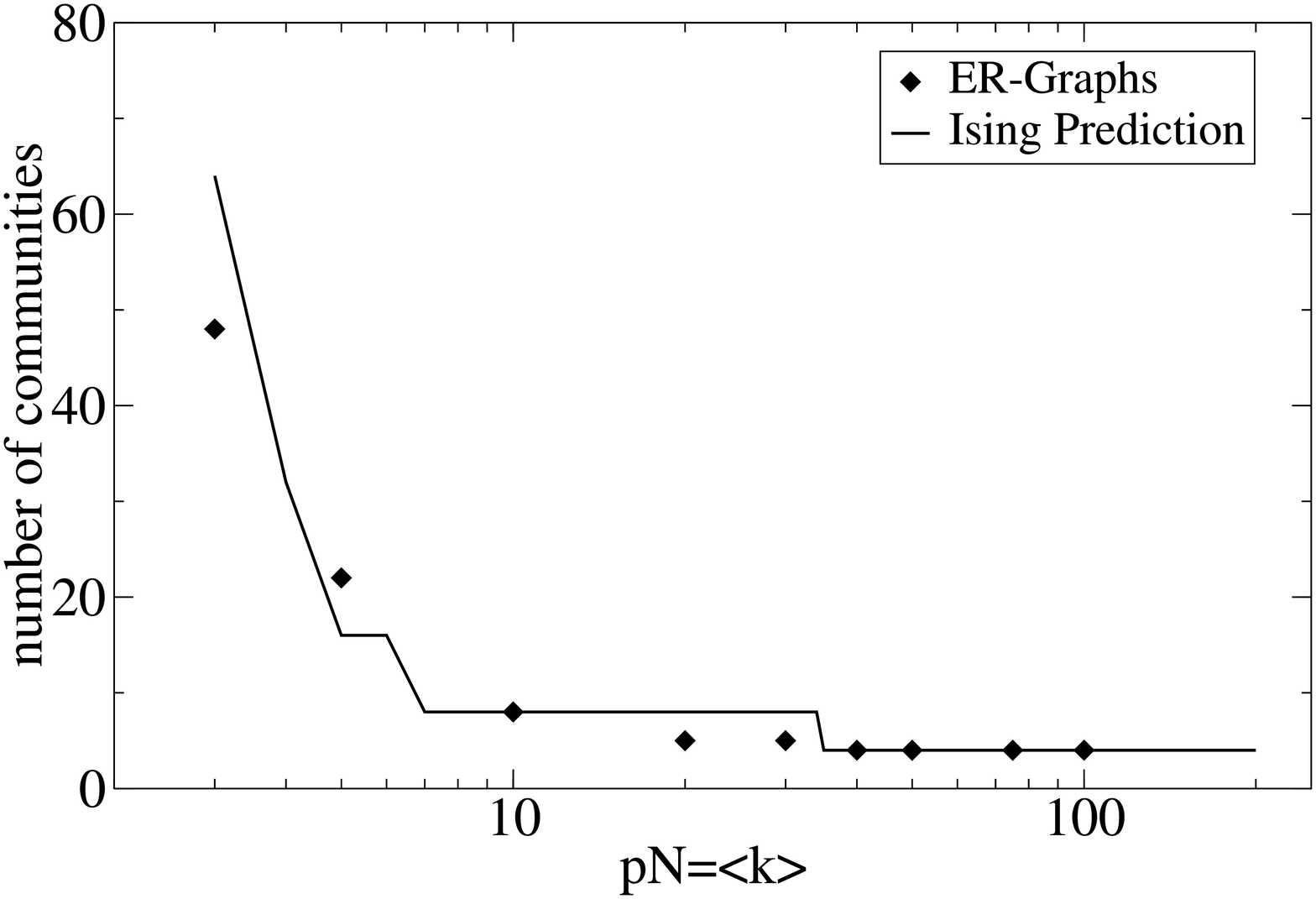}
\caption{Modularity and the number of communities in ER-Graphs. Shown are the values determined from clustering random graphs with $N=10,000$ nodes and the expectation values calculated from using a Potts-Model (\ref{ModularityEstimation}) or an Ising-Model (\ref{IsingEstimation}) recursively.  }
\label{RandomGraphs}
\end{figure}

We see that maximum modularity is obtained at $q=5$, though the value of $U(q)/q$ for $q=4$ is not much different from it. This qualitative behavior of dense random graphs tending to cluster into only a few large communities is confirmed by our numerical experiments. 
Using the largest value of Table \ref{PottsGS}, we finally arrive at an expression for the modularity that we can expect in any ER random graph with average degree $\langle k\rangle=pN$: 
\begin{equation}
Q=0.97\sqrt{\frac{1-p}{pN}}.
\label{ModularityEstimation}
\end{equation}
Figure \ref{RandomGraphs} shows the comparison of equation (\ref{ModularityEstimation}) and experiments where we have numerically maximized the modularity in random graphs with $N=10,000$ nodes and varying connectivity $\langle k\rangle$ using a simulated annealing approach as described in an earlier section.

The above approximation using a Potts spin glass, however, cannot explain the number of communities found experimentally in random graphs of varying connectivity since it always assumes $5$ communities. Therefore, we try to approximate the ground state of a $q$-state Potts model by recursively bi-partitioning the network and continuing as long as the modularity increases. For every bi-partition we use the expression of the cut-size as a function of the number of $N$ nodes and average degree $\langle k\rangle=pN$ given by Fu and Anderson \cite{FuAnd}:
\begin{equation}
 	\min_{\{\sigma_i\}}\left[\mathcal{C}(\{N/2\})\right]=\frac{M}{2}\left[1-c\sqrt{\frac{1-p}{pN}}\right].
\end{equation}
The constant $c$ corresponds to $U(2)$ and is given by $c=1.5266\pm0.0002$ \cite{FuAnd}. After every partition, the number of links connecting to nodes in the same part and to nodes in the rest of the network is given by:
 \begin{equation}
 	\langle k_{in}\rangle=\frac{pN+c\sqrt{pN(1-p)}}{2}\hspace{0.5cm}\mbox{and}\hspace{0.5cm}\langle k_{out}\rangle=\frac{pN-c\sqrt{pN(1-p)}}{2}.
	\label{KinKout}
 \end{equation}
After $b$ successive recursive partitions, we arrive at a modularity of 
 \begin{equation}
 	Q(b)=\frac{2^b-1}{2^b}-\frac{1}{\langle k\rangle}\sum_{t=1}^b\langle k_{out,t}\rangle
	\label{IsingEstimation}
 \end{equation}
where $\langle k\rangle$ is the average degree in the total network and $\langle k_{out,t}\rangle$ is the average number of external links a node gains after partition number $t$ calculated from (\ref{KinKout}). 

Though equation (\ref{IsingEstimation}) only allows numbers of communities that are powers of 2, the agreement with the experimental data is surprisingly good as Figure \ref{RandomGraphs} shows. Also, the number of communities is predicted almost perfectly by (\ref{IsingEstimation}) as shown in Figure \ref{RandomGraphs}.

With the expression (\ref{ModularityEstimation}) and (\ref{IsingEstimation}), we are adequately able to calculate expectation values of $Q$ for random graphs which can be used in the assessment of the statistical significance of the modularity in real world networks. We have shown, that random graphs may exhibit considerable values of modularity even without any built-in group structure. Significant community structure can hence only be attributed to graphs with values of modularity higher than those calculated for random graphs. The sparser a graph, the higher the modularity of its randomized equivalent. It is hence especially difficult to detect true modularity in sparse graphs. Also, the sparser a graph, the more modules it will show, while dense random graphs tend to cluster into only a hand full of communities. 

\section{Theoretical Limits of Community Detection}
\begin{figure}
\includegraphics[width=10cm]{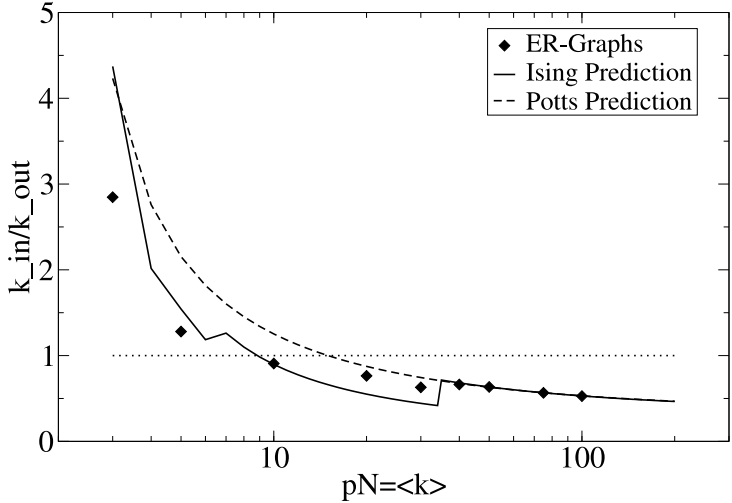}
\caption{Ratio of internal links to external links $k_{in}/k_{out}$ in the ground state of the Hamiltonian. Shown are the experimental values from clustering random graphs with $N=10,000$ nodes and the expectation values calculated from using a Potts-Model (\ref{ModularityEstimation}) or an Ising-Model (\ref{IsingEstimation}) recursively. The dotted line represents the Radicchi \etal definition of community in ``strong sense'' \cite{Radicchi}. Note that sparse graphs will, on average, always exhibit such communities, while dense graphs will not, even though their modularity may be well above the expectation value for an equivalent random graph.}
\label{kRatio}
\end{figure}

With the results of the last section we are now in the position to explain Figure \ref{Benchmark} and to give a limit to which extent a designed community structure in a network can be recovered. As we have seen, for any random network we can find an assignment of spins in communities that leads to a modularity $Q>0$. For our computer-generated test networks with $\langle k\rangle=16$ we have a value of $p=\langle k\rangle/(N-1)=0.126$ and expect a value of $Q=0.227$ according to (\ref{ModularityEstimation}) and $Q=0.262$ according to (\ref{IsingEstimation}). The modularity of the community structure built in by design is given by:
\begin{equation}
Q(\langle k_{in}\rangle)=\frac{\langle k_{in}\rangle}{\langle k\rangle}-\frac{1}{4}.
\end{equation} 
Hence, below $\langle k_{in}\rangle=8$, we have a designed modularity that is smaller than what can be expected from a random network of the same connectivity! This means that the minimum in the energy landscape corresponding to the community structure that we design is less deep than those that one can find in the energy landscape defined by any network. It must be understood that in the search for the built in community structure, we are competing with those community structures that arise from the fact that we are optimizing for a particular quantity in a very large search space. In other words, any network possesses a community structure that exhibits a modularity at least as large as that of a completely random network. If a community structure is to be recovered reliably, it must be sufficiently pronounced in order to win the comparison with the structures arising in random networks. In the case of the test networks employed here, there must be more than $\approx 8$ intra-community links per node.  Figure \ref{kRatio} again exemplifies this. We see that random networks with $\langle k\rangle=16$ are expected to show a ratio of internal and external links $k_{in}/k_{out}\approx 1$. Networks which are considerably sparser have a higher ratio while denser networks have a much smaller ratio. This means that in dense networks, we can recover designed community structure down to relatively smaller $\langle k_{in}\rangle$. Consider for example large test networks with $\langle k\rangle=100$ with 4 built-in communities. For such networks, we expect a modularity of $Q\approx 0.1$ and hence the critical value of intra-community links to which the community structure could reliably be estimated would be $\langle k_{in}\rangle_c=35$ which is much smaller in relative comparison to the average degree in the network. 

This also means, that the point at which we cannot distinguish between a random and a modular network is not defined by $p_{in}=p_{out}=p$ for the internal and external link densities as one may have intuitively expected. Rather, it is determined by the ratio of $\langle k_{in}\rangle/(\langle k\rangle-\langle k_{in}\rangle)$ in the ground state of a random network and depends on the connectivity of the network $\langle k\rangle$. 

Finally, from Figure \ref{kRatio} we observe that sparse random graphs all show communities in the strong sense of Radicchi \etal \cite{Radicchi}. Further, it is very difficult to find communities in the strong sense in dense graphs, even though they may exhibit a modularity well above that of a random graph.  

\section{Conclusion}
Starting from a simple principle, we have shown how the problem of community detection can be mapped onto finding the ground state of an infinite range spin glass. The quality function of the clustering is identified as the ground state energy of this spin glass.  Benchmarks show the good performance of algorithms based on this mapping. The network modularity $Q$ defined by Girvan and Newman is identified as a special case of this approach. The comparison with appropriate random graphs allows the assessment of the statistical significance of community structures found in real world networks. Expectation values for the modularity of Erd\H{o}s-R\'{e}nyi random graphs were given. The theoretical limits of community detection were addressed and we found that only those community structures can be recovered reliably that lead to modularities larger than the expectation values of random graphs.  

\section{Acknowledgements}
The authors would like to thank Stefan Braunewell, Michele Leone, Ionas Erb and Andreas Engel for many helpful hints and discussions.


\bibliography{../BibTex_Citations}

\end{document}